\documentclass[letterpaper,english,prl,twocolumn,superscriptaddress,groupedaddress]{revtex4}
\usepackage[T1]{fontenc}
\usepackage[latin9]{inputenc}
\usepackage{color}
\usepackage{babel}
\usepackage{amstext}
\usepackage{amssymb}
\usepackage{graphicx}
\usepackage{esint}
\usepackage[unicode=true,pdfusetitle,
 bookmarks=true,bookmarksnumbered=false,bookmarksopen=false,
 breaklinks=false,pdfborder={0 0 1},backref=false,colorlinks=false]
 {hyperref}

\makeatletter

\pdfpageheight\paperheight
\pdfpagewidth\paperwidth

\DeclareRobustCommand{\greektext}{%
  \fontencoding{LGR}\selectfont\def\encodingdefault{LGR}}
\DeclareRobustCommand{\textgreek}[1]{\leavevmode{\greektext #1}}
\DeclareFontEncoding{LGR}{}{}
\DeclareTextSymbol{\~}{LGR}{126}
\providecommand{\tabularnewline}{\\}

\@ifundefined{textcolor}{}
{%
 \definecolor{BLACK}{gray}{0}
 \definecolor{WHITE}{gray}{1}
 \definecolor{RED}{rgb}{1,0,0}
 \definecolor{GREEN}{rgb}{0,1,0}
 \definecolor{BLUE}{rgb}{0,0,1}
 \definecolor{CYAN}{cmyk}{1,0,0,0}
 \definecolor{MAGENTA}{cmyk}{0,1,0,0}
 \definecolor{YELLOW}{cmyk}{0,0,1,0}
}

\makeatother

\begin{document}

\title{Full-field quantum correlations of spatially entangled photons}

\author{V. D. Salakhutdinov}

\address{Huygens Laboratory, Leiden University, P.O. Box 9504, 2300 RA Leiden,
The Netherlands}

\author{E. R. Eliel}

\address{Huygens Laboratory, Leiden University, P.O. Box 9504, 2300 RA Leiden,
The Netherlands}

\author{W. Löffler}

\address{Huygens Laboratory, Leiden University, P.O. Box 9504, 2300 RA Leiden,
The Netherlands}

\address{loeffler@physics.leidenuniv.nl}
\begin{abstract}
Spatially entangled twin photons allow the study of high-dimensional
entanglement, and the Laguerre-Gauss modes are the most commonly used
basis to discretize the single photon mode spaces. In this basis,
to date only the azimuthal degree of freedom has been investigated
experimentally due to its fundamental and experimental simplicity.
We show that the full spatial entanglement is indeed accessible experimentally,
i.e., we have found practicable radial detection modes with negligible
cross correlations. This allows us to demonstrate hybrid azimuthal
-- radial quantum correlations in a Hilbert space with more than 100
dimensions per photon.
\end{abstract}
\maketitle

High-dimensional entangled photons are of great interest in various
areas in quantum information, such as they promise high-density encoding
of quantum information \cite{bennett1992,bechmannpasquinucci2000},
are more robust against noise and eavesdroppers due to stronger non-classical
correlations \cite{kaszlikowski2000}, and in general, present a unique
model system for the study of high-dimensional entanglement in nature.
Entanglement in the photon's spatial degrees of freedom is a candidate
for this, and it can readily be obtained in the lab by spontaneous
parametric downconversion (SPDC) of an intense laser beam. To explore
this high-dimensional Hilbert space, we need to discretize this initially
continuous space; due to the paraxial nature of experiments, this
is usually done using a complete and orthogonal basis of transverse
optical modes. In quantum information with entangled particles, it
is crucial that the bipartite state shows perfect correlation (or
anticorrelation) in the used quantum numbers. Paraxial optical modes
are also required for the implementation of quantum cryptography:
they are propagation-invariant and superpositions thereof are basically
stable. Traditionally, a Gaussian basis is employed, in particular
the Laguerre-Gaussian modes $LG_{p}^{\ell}$ with mode indices $\ell$
and $p$ proved to be very convenient. These modes factor in an azimuthal
phase-only part $u_{az}^{\ell}(\phi)=e^{i\ell\phi}$ and a radial
part $u_{rad}^{p,\ell}(r)$, in which the azimuthal part gives rise
to the photon orbital angular momentum (OAM) of $\ell\hbar$ \cite{allen1992}.
This azimuthal part, or OAM entanglement, has been subject to a decade
of numerous very successful experiments (see, e.g., ref. \cite{mair2001,dada2011}),
which is well founded by the fact that most experimental setups exhibit
rotation symmetry around the optical axis. 

\begin{figure}
\includegraphics[width=1\columnwidth]{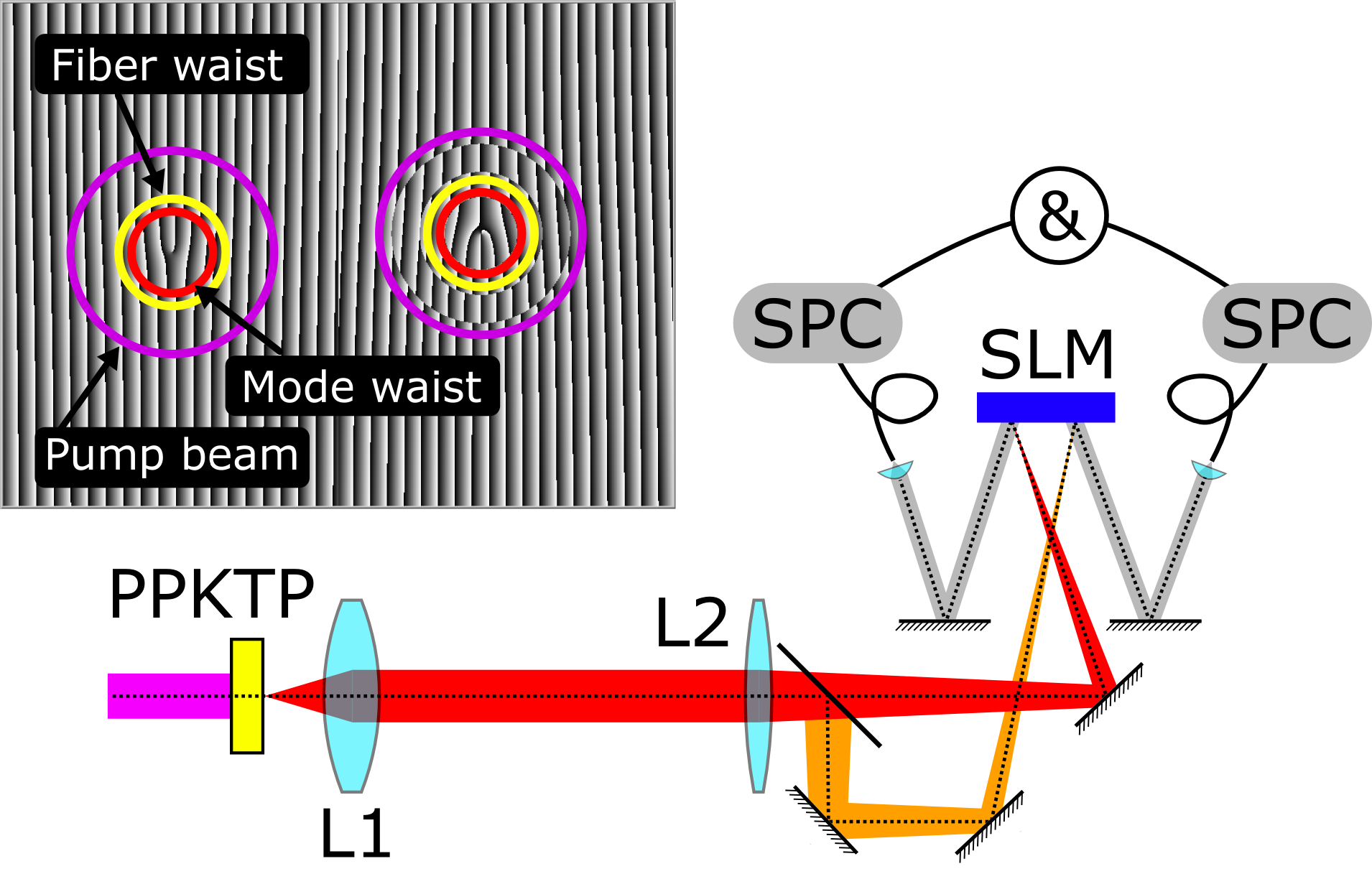}

\caption{Schematic experimental setup. The spatially entangled photons are
produced in the nonlinear crystal, whose surface is imaged with $7.5\times$
magnification ($f_{L1}=10$ cm, $f_{L2}=75$~cm) onto the spatial
light modulator (display size $16\times12$~mm$^{2}$, $800\times600$
pixel). This device is programmed to perform the phase modulation
required to transform the detection mode into the fundamental mode.
The far-field of the SLM surface is imaged ($10\times$, 0.2 NA objectives)
onto the single-mode fiber which is connected to a single photon counter.
Simultaneous detection events from an entangled photon pair are selected
in a coincidence time window of 2.3~ns. The inset shows an exemplary
SLM phase pattern ($\ell_{s}=1,\, p_{s}=1;\,\ell_{i}=-2,\, p_{i}=3)$
and superimposed the magnification-corrected waist of the pump beam
(purple), the waist of the detection singlemode fiber (1275~mm, yellow),
and the detection mode waist (red) $w=1000$~\textgreek{m}m.}
\end{figure}
The amount of entanglement present in the spatial photon pairs can
be characterized by the \emph{average} number of entangled optical
modes, the Schmidt number $K$ \cite{law2004,torres2003}. This Schmidt
number $K=1/\sum_{k}\lambda_{k}^{2}$ is obtained from the eigenvalues
(relative weights) $\lambda_{k}$ of the Schmidt decomposition \cite{ekert1995}
of the two-photon field $|\Psi\rangle=\sum_{k}\sqrt{\lambda_{k}}|u_{k}\rangle_{s}|u_{k}\rangle_{i}$,
where the $|u_{k}\rangle_{s,i}$ are the Schmidt eigenmodes for the
signal or idler photon. Although the Schmidt modes have to be calculated
numerically in the general case, the Schmidt number $K$ can be approximated
as $K=\frac{1}{4}(b\sigma+\frac{1}{b\sigma})^{2}$, where $b^{-1}$
(with $b^{2}=L\lambda_{p}/8\pi$) is the phase-matching width, and
$\sigma$ is the pump beam waist \cite{law2004}. For our experimental
parameters (crystal length $L=2$~mm, pump beam waist $w_{p}=325$~\textgreek{m}m,
pump beam wavelength $\lambda_{p}=413$~nm), this number is very
large: $K\approx350$. However, if only the azimuthal degree of freedom
is employed (i.e., taking $p=0$ \cite{osorio2008}) this number is
significantly lower. We can write the two-photon entangled state as
$|\psi\rangle=\sum_{\ell=-\infty}^{+\infty}\sqrt{\lambda_{\ell}}|\ell,p=0\rangle_{s}|-\ell,p=0\rangle_{i}$,
where $|\ell,p\rangle$ is a photon with OAM $\ell\hbar$ and radial
quantum number $p$, and the (azimuthal) Schmidt number becomes $K_{az}=1/\sum_{\ell}\lambda_{\ell}^{2}$.
For large $K$, this can be approximated as $K_{az}\approx2\sqrt{K}$
\cite{vanexter2006,vanexter2007}. Direct experimental determination
of this number has been shown only recently \cite{dilorenzopires2010}.
For our case, this number is $K_{az}\approx37$, which is obviously
much lower than the total number of entangled modes. The ``missing''
entanglement becomes accessible if also the radial modes are taken
into account. There, we find a \emph{radial} Schmidt number (for only
one azimuthal mode, e.g., for $\ell=0$) $K_{rad}\approx\sqrt{K}$,
which in our example is $K_{rad}\approx18$. The vast majority of
the entangled modes are radial-azimuthal cross-correlated modes \cite{straupe2011}. 

Subject to experimental feasibility, the radial part of the Laguerre-Gauss
entangled modes is an entanglement resource on equal footing; however,
only recently it has been investigated in detail theoretically \cite{miatto2011}.
The LG mode functions factor as $LG_{p}^{\ell}(r,\phi)=C\cdot u_{az}^{\ell}(\phi)\cdot u_{rad}^{p,\ell}(r)$;
the azimuthal part is fully orthogonal in $\ell$ and independent
on the experimental choice of the detection mode waist, and the entangled
photons are perfectly anti-correlated in $\ell$: OAM is conserved
in SPDC (In Fig.~S1 of the supplementary information we show that
this statement holds also for higher-order radial modes). Therefore,
the azimuthal modes are automatically Schmidt modes. It turns out
that in contrast to this, the radial modes do not necessarily represent
Schmidt modes \cite{law2004} and we expect to find non-zero quantum
correlations of detected modes with different $p$. However, for a
proper combination of pump-beam, detection-mode and phase matching
waist $w_{s,i}^{\ast}=\sqrt{4b/\sigma}$ \cite{law2004,vanexter2006},
also the radial LG modes are Schmidt modes and the cross-correlations
for $p_{s}\neq p_{i}$ disappear. In our case, we obtain $w_{s,i}^{\ast}=37$~\textgreek{m}m.
If we neglect phase matching \cite{miatto2011}, $w_{s,i}^{\ast}\rightarrow0$.

Also experimentally, investigation of the radial correlations turns
out to be more challenging: The spatial correlations are traditionally
investigated using a mode converter (spatial light modulator or spiral
phase plate in the case of azimuthal correlations), to transform a
certain optical mode into the fundamental Gaussian, which in turn
can be tested by sending the photon into a single-mode fiber. This
works very well for the azimuthal modes, but for radial modes, complications
occur: for instance, the finite acceptance angle of the SMF becomes
problematic, and there are no perfect spatial modulators which allow
control over amplitude \emph{and} phase simultaneously (the orthogonality
of $p$-modes requires amplitude-sensitive detection), and careful
choices for the detection mode waist and the fiber collimator have
to be taken \cite{law2004}. We show here that despite of these complications,
and under the right conditions, the relation between the analyzer
modes and the pump field becomes nicely visible, and a properly correlated
mode basis can be obtained. 

\emph{\medskip{}
}

\emph{Experiment. }We generate the spatially entangled photon pairs
by collinear SPDC in a PPKTP crystal (length $L$~=~2~mm) of a
$LG_{0}^{0}$ laser beam ($\mathsf{Kr^{+}}$, $\lambda$~=~413~nm,
beam waist at crystal $w_{p}$~=~325~\textgreek{m}m, 50~mW power). 

\begin{figure}
\includegraphics[width=1\columnwidth]{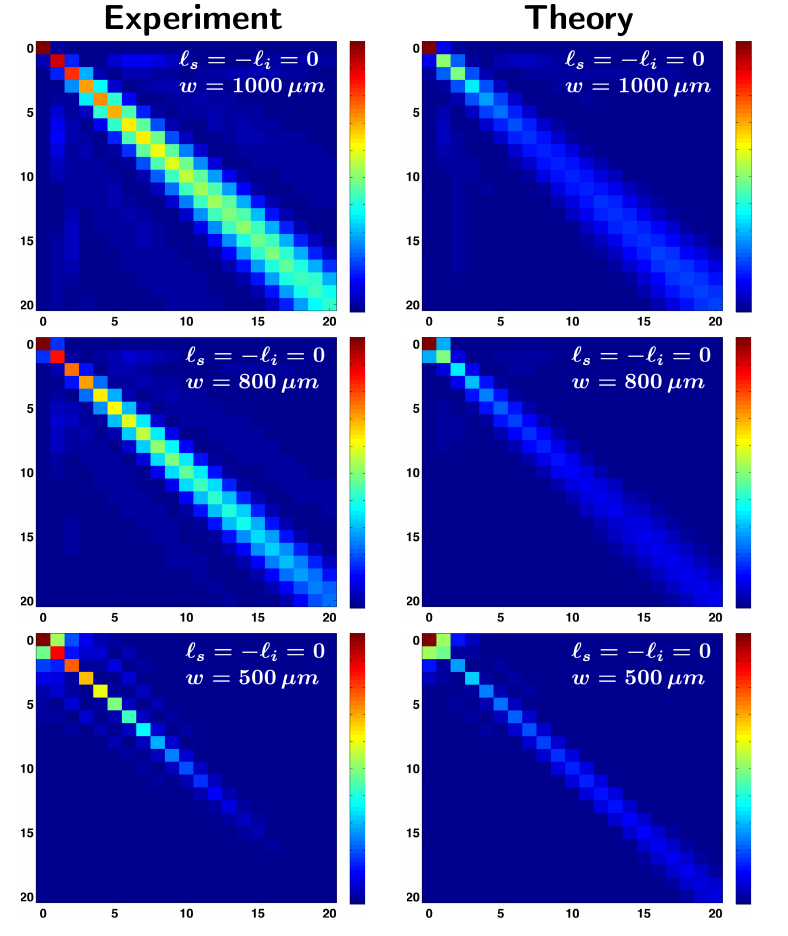}\caption{Quantum correlations between radial modes with different $p$ (for
$\ell_{s}=\ell_{i}=0$): Shown are the normalized \textcolor{magenta}{}(divided
by maximum) coincidence count rates (color coded) as a function of
the radial mode numbers $p_{s}$ (horizontal axis) and $p_{i}$ (vertical
axis) of the detection modes. Different rows depict results for different
detection mode waists as indicated. Left column: experimental data,
right column: theoretical prediction. It is clearly visible that the
smaller the detection mode waist gets, the smaller the off-diagonal
counts will be. This is a sign that we approach the Schmidt basis
for $\gamma\rightarrow\gamma^{\ast}$. The detection mode waists corresponds
to waist ratios of (from top to bottom) $\gamma=$~2.4; 3; 4.9, see
\cite{miatto2011}.}
\end{figure}
As sketched in Fig.~1, we image the crystal surface with $7.5\times$
magnification using a telescope onto the SLM surface. The SLM is used
under an incident angle of 10 or 5 degrees; this allows us to use
a single SLM for both signal and idler photon. The SLM is corrected
for phase flatness, and we operate not in direct phase modulation,
but use a blaze towards 2 mrad to further lessen the influence of
phase errors. The far field of the SLM surface is imaged onto the
single mode fiber using $10\times$ objectives, with a detection-fiber
mode waist at the SLM of 1275~\textgreek{m}m. The fibers are connected
to single-photon counters and we post-select entangled photon pairs
by coincidence detection (time window 2.3~ns). Since the crystal
surface is \emph{imaged} onto the SLM, it is sufficient to discuss
the situation there. The inset of Fig.~1 shows the resulting waists
of the pump beam, the detection mode, and the detection single-mode
fiber; with exemplary phase patterns for two different settings of
the detection mode quantum numbers. The choice of waists depends on
(i) the desired ratio $\gamma=w_{p}/w_{s,i}$ which determines the
orthogonality and overlap with the Schmidt modes (where the ideal
ratio is $\gamma^{\ast}=8.8$), (ii) the maximum mode order which
should be detected, this is connected to the singlemode detection
fiber mode waist, and (iii) the number of entangled modes required.
Our choice of waists is optimized for radial and azimuthal mode numbers
up to about 10.

Our SLM-based mode detectors can not project upon perfect LG modes,
because the amplitude cannot be modulated (this is not possible with
conventional SLMs %
\footnote{Amplitude-shaped holograms (as used in, e.g., \cite{dada2011}) are
not suitable for small-angle holograms like the ones used here, they
do not produce phase-correct fields in the far field of the first
diffraction order.%
}). This can lead to $p$-non-orthogonal detection fields $u_{p}^{\ell}$
because $\int_{0}^{\infty}\arg[LG_{p_{1}}^{\ell}(r)\, LG_{p_{2}}^{\ell}(r)]\neq\delta_{p_{1},p_{2}}$,
and one would anticipate that cross-correlations will always appear;
our results below show that this is not always true, and that careful
adjustment of the detection mode waist allows detection of radially
entangled modes with negligible cross-correlations. Basically, optical
diffraction couples phase and amplitude, which helps to obtain amplitude-sensitive
detection.

For theoretical calculation of the expected coincidence count rate,
we apply Klysko's picture of advanced waves \cite{klyshko1988}. The
detection field (in the near field of the SLM) is determined by the
Gaussian amplitude of the single-mode detection fiber, and the phase
as defined by the SLM: $u_{xtal}^{p,\ell}=\exp\left[i\,\text{arg}\left[LG_{p}^{\ell}\right]-r^{2}/w_{SMF}^{2}\right]$.
We then decompose this field in terms of LG modes: $u_{xtal}^{p,\ell}=\sum_{p'}\alpha_{p',\ell}\cdot LG_{p'}^{\ell}$.
This expansion contains very high-order $p$-components due to the
phase jumps at the zeros of the LG polynomial with finite intensity.
These singularities automatically disappear while weighting the modes
with their relative weight as produced in SPDC $C_{p,p}^{\ell,-\ell}$
(Eq.~20 in \cite{miatto2011}). This results in $\alpha_{p',\ell}'=2\pi\sqrt{C_{p',p'}^{\ell,-\ell}}\int\mathrm{d}r\, r\, LG_{p'}^{\ell}(r)\cdot u_{xtal}^{p,\ell}$,
which allows us to calculate the effective detection field at the
crystal $u_{det}^{p,\ell}=\sum_{p'}\alpha_{p',\ell}\cdot LG_{p'}^{\ell}$.
To obtain the coincidence amplitude, we can then simply calculate
the overlap of the two individual detection fields in the signal and
idler path: $C'=\int\mathrm{d^{2}}r\, u_{det}^{p_{s},\ell_{s}}\cdot u_{det}^{p_{i},\ell_{i}}$.
Finally, the experimentally observable coincidence count rate is $\Gamma=|C'|^{2}$.
We use here the the thin-crystal limit ($L\rightarrow0$), which implies
that phase-matching effects are neglected (experimentally, we are
close to perfect phase matching).

\begin{figure}
\includegraphics[width=1\columnwidth]{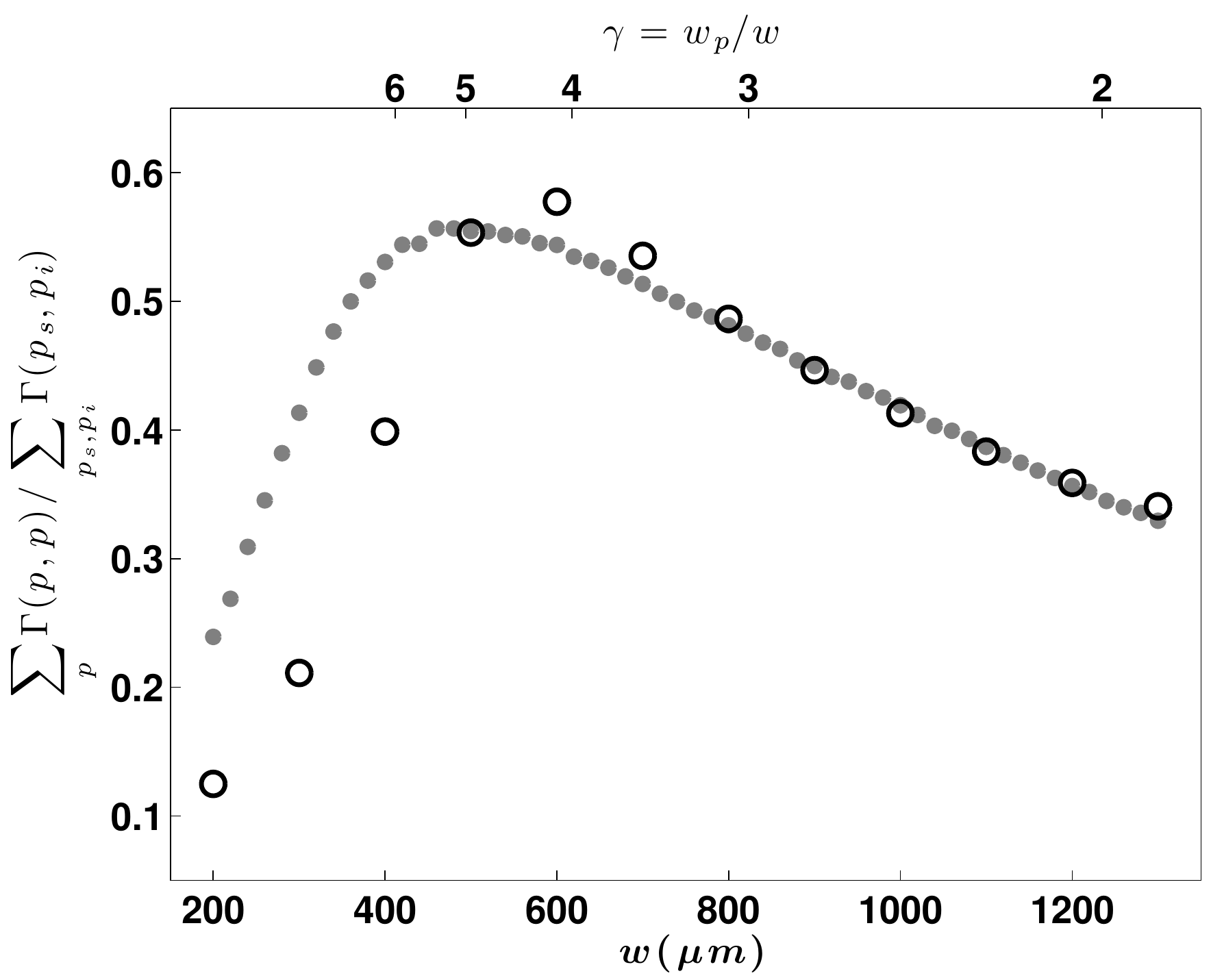}\caption{The influence of detection mode waist: The quantity on the $y$-axis
is a measure of how sharply the radial quantum correlations peak around
$p_{s}=p_{i}$. For ideal LG mode detectors, we have $W\rightarrow1$
for the waist ratio $\gamma\rightarrow\gamma^{\ast}$. For small waists
($w<500\,\mu\mathrm{m}$), pixilation effects are non-negligible.
For larger waists, the agreement between experiment (circles) and
theory (grey dots) is very good. The top axis indicates the ratio
$\gamma$ of the pump beam waist $w_{p}$ to the detection mode waist
$w=w_{s,i}$ of the signal and idler photon.}
\end{figure}

\emph{\medskip{}
}

\emph{Radial-mode correlations. }Fig.~2 shows the quantum correlations
of purely radial modes ($\ell_{s}=\ell_{i}=0$) of downconverted photons.
We clearly observe, as we decrease the detection mode waist, that
the off-diagonal elements in the correlation matrix decrease. This
is expected, as mentioned above, the radial cross-correlations disappear
for $w_{s,i}\rightarrow w_{s,i}^{\ast}$. To within our experimental
accuracy, we also reproduce the theoretical results of Miatto et al.
\cite{miatto2011} very well. Even minute details of the experimental
data are reproduced qualitatively well by our model (Fig.~2); this
suggests that our modelling approach of expanding the detection field
in terms of the LG modes provided by the SPDC light, is a sound choice.
For the case of 500~\textgreek{m}m mode waist, we estimate the (radial)
Schmidt number to be 10.4 (experiment) and 11.2 (theory).  This is
less than the expectation mentioned above ($K_{rad}=18$), however,
SLM pixilation becomes relevant at such small mode waists. To investigate
this, we determine a measure of the cross correlations, or the width
of the diagonal (Fig.~2) around $p_{s}=p_{i}$: $W=\sum_{p}\Gamma(p_{s}=p_{i}=p)/\sum_{p_{s},p_{i}}\Gamma(p_{s},p_{i})$.
For perfectly orthogonal modes, $W$ should be unity. Fig.~3 shows
$W$ as a function of the beam waist, comparing our theoretical simulation
with experimental data, again we find good agreement. This dependency
of the cross-correlations on beam waist ratio persists also for higher
azimuthal modes ($\ell_{s}=-\ell_{i}$), as shown in Fig.~S2 of the
supplementary information. We observe that for a detection mode waist
smaller than 500~\textgreek{m}m, the ``orthogonality'' $W$ decreases
again, this (and the fact that the theoretical curve does not reach
unity) is a consequence of SLM pixilation: 500~\textgreek{m}m waist
corresponds to $\sim25$ pixel of the SLM. Our results also demonstrate
that the apprehension of Miatto et al. \cite{miatto2011}, that the
experimentally accessible mode waist ratios $\gamma$ are too small,
therefore leading to strong cross-correlation in $p$-space, which
would imply that radial modes are not useful for quantum information,
is overcautious: We can adjust the mode waists that cross-correlations
become negligible.

\begin{figure}
\includegraphics[width=1\columnwidth]{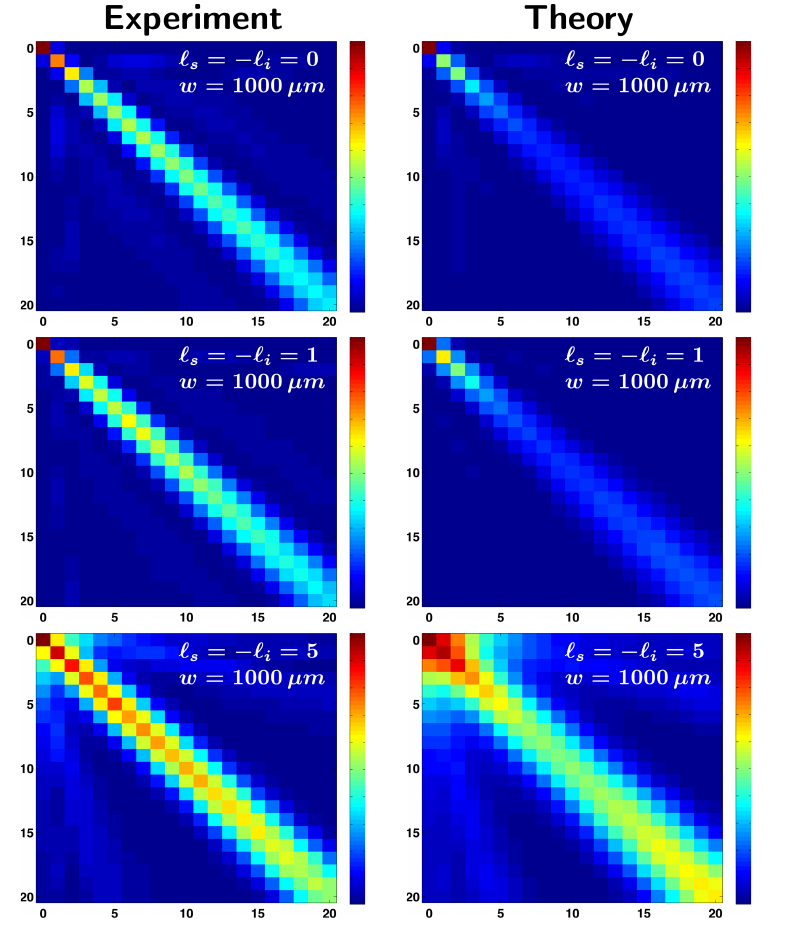}\caption{Quantum correlations between radial modes with different $p$, for
given $\ell$ (rows) at a fixed waist ratio $\gamma=2.4$. The left
column shows experimental data, the right column the theoretical prediction.
The number of single-particle modes involved in these experiments
is approximately 100.}
\end{figure}

\emph{\medskip{}
}

\emph{Radial-azimuthal correlations. }Finally, we address the question
if we can make use of the full azimuthal-radial Hilbert space \emph{experimentally}.
We find, independent on the radial mode index $p$, the anticorrelation
condition $\ell_{s}=-\ell_{i}$, or OAM conservation, is preserved
for high values (up to $\approx20$) of $p$ and $\ell$ (see Fig.~S1).
Fig.~4 shows $p$-correlations for a fixed detection mode waist of
1000~\textgreek{m}m, as in Fig.~2, but this time for different choices
of $\ell_{s}=-\ell_{i}\equiv\ell$. Compared to $\ell=0$, we observe
negligible increase of the cross correlations for $p_{s}\neq p_{i}$,
which is very encouraging, because this suggests that very high dimensional
Hilbert spaces become accessible. Additionally, the $p_{s}=p_{i}$
correlations get more evenly distributed, in agreement with theoretical
predictions \cite{miatto2011}, which also increases the number of
usable modes. Our results in Fig.~4 show two-photon correlations
in an approximately $100\times100$ dimensional Hilbert space. 

\medskip{}

\emph{Conclusions. }In conclusion, we have shown first experiments
with high-dimensionally spatially entangled photons in the full Laguerre-Gauss-like
basis. We analyze the entangled photons in the complete transverse
basis involving azimuthal and radial correlations; this goes a step
forward beyond the conventionally used azimuthal degree of freedom,
or orbital angular momentum entanglement. We find that the radial
degree of freedom is indeed a useful entanglement resource, if care
is taken: Our experiments and the theoretical model show that the
choice of detection mode waist is crucial and has to be taken into
account carefully; we are able to demonstrate the transition to a
detection basis where cross-correlations disappear, effectively a
transition to a quasi-Schmidt basis. An important next step will be
confirmation and quantification of the ``hybrid'' azimuthal--radial
mode entanglement, which is beyond the scope of this paper. If radial
modes in spatial entanglement are accessible, the number of useful
entangled modes is roughly squared compared to the OAM case, this
quadratic increase in the usable Schmidt number could stimulate new
experiments like detection-loophole free \cite{vertesi2010} Bell
tests. The higher entanglement density per mode area will also enable
higher channel capacities in systems where the spatial extent is relevant:
For the transport of spatially entangled photons through optical fibers
\cite{loeffler2011}, and also through turbulent atmosphere \cite{pors2011}.

We gratefully acknowledge fruitful discussions with M. van Exter,
G. Nienhuis, F. Miatto, and H. Woerdman and financial support by NWO,
the Gorter Fonds, and the European Union Commission within the 7th
Framework project \# 255914 PHORBITECH.

\cleardoublepage{}

\begin{widetext} 

\begin{center}
\textbf{\large Supplementary information }
\par\end{center}{\large \par}

\begin{tabular}{ccc}
$p_{s}=p_{i}=0$ & $p_{s}=p_{i}=5$ & $p_{s}=p_{i}=10$\tabularnewline
\includegraphics[width=0.3\columnwidth]{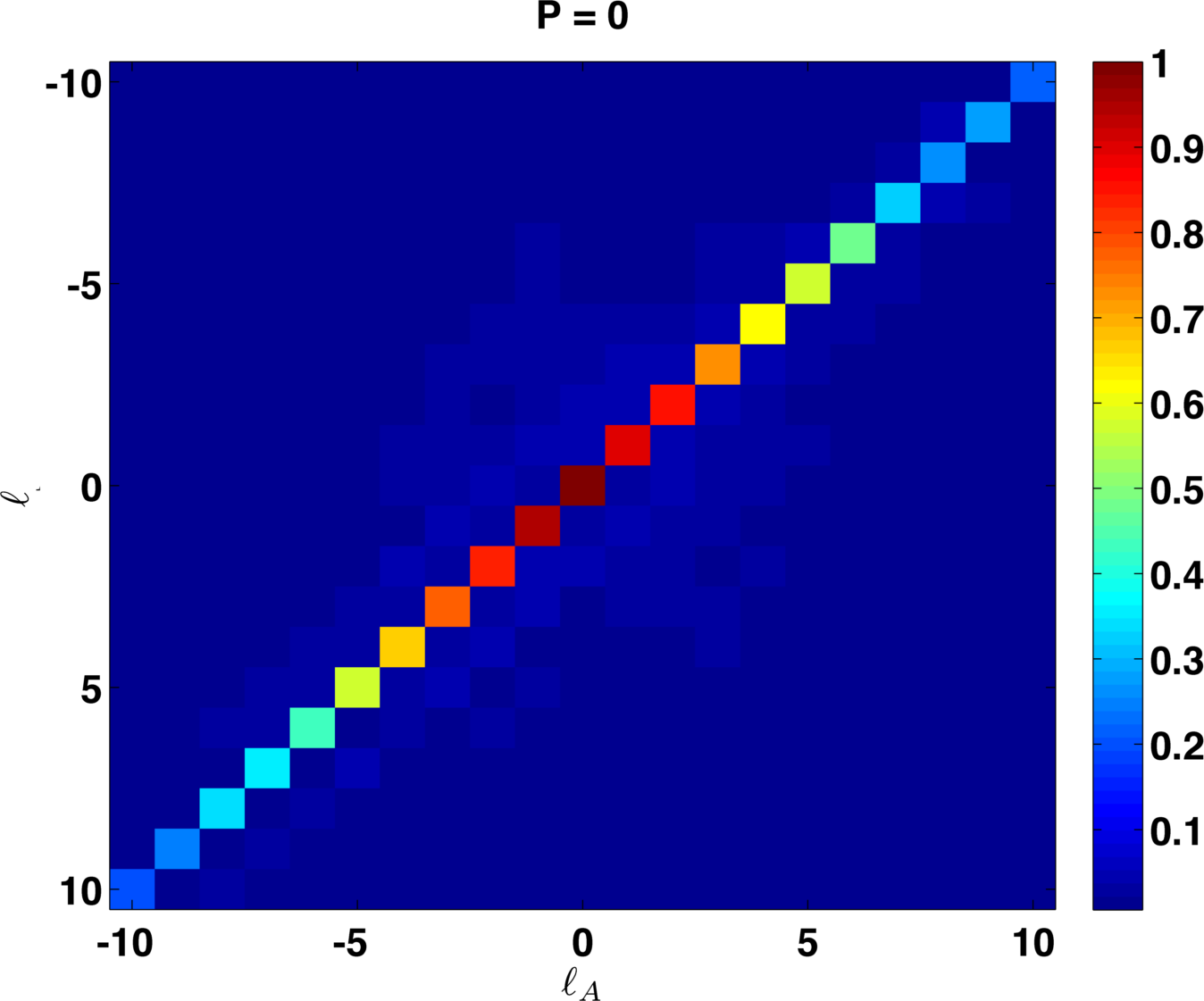} & \includegraphics[width=0.3\columnwidth]{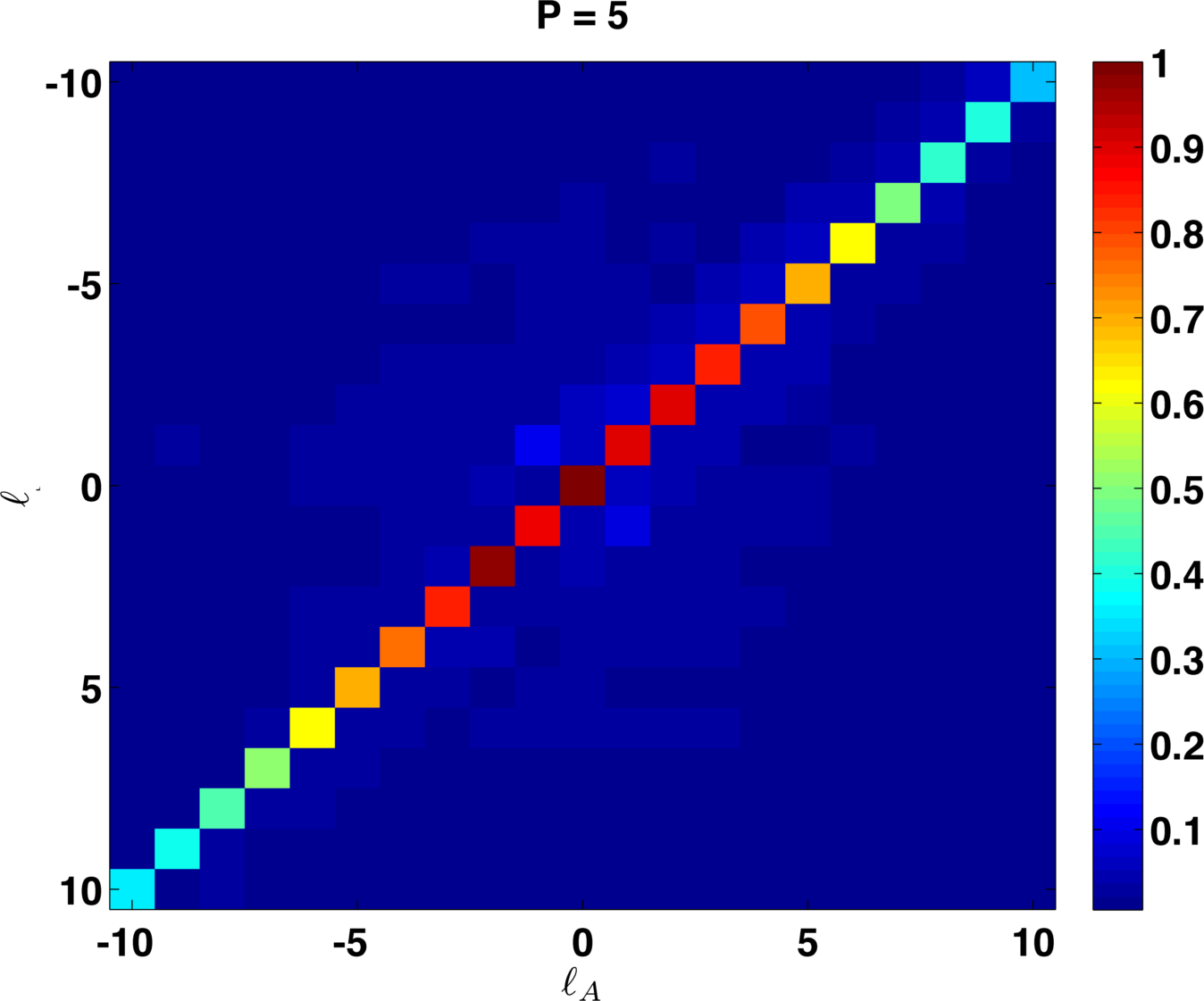} & \includegraphics[width=0.3\columnwidth]{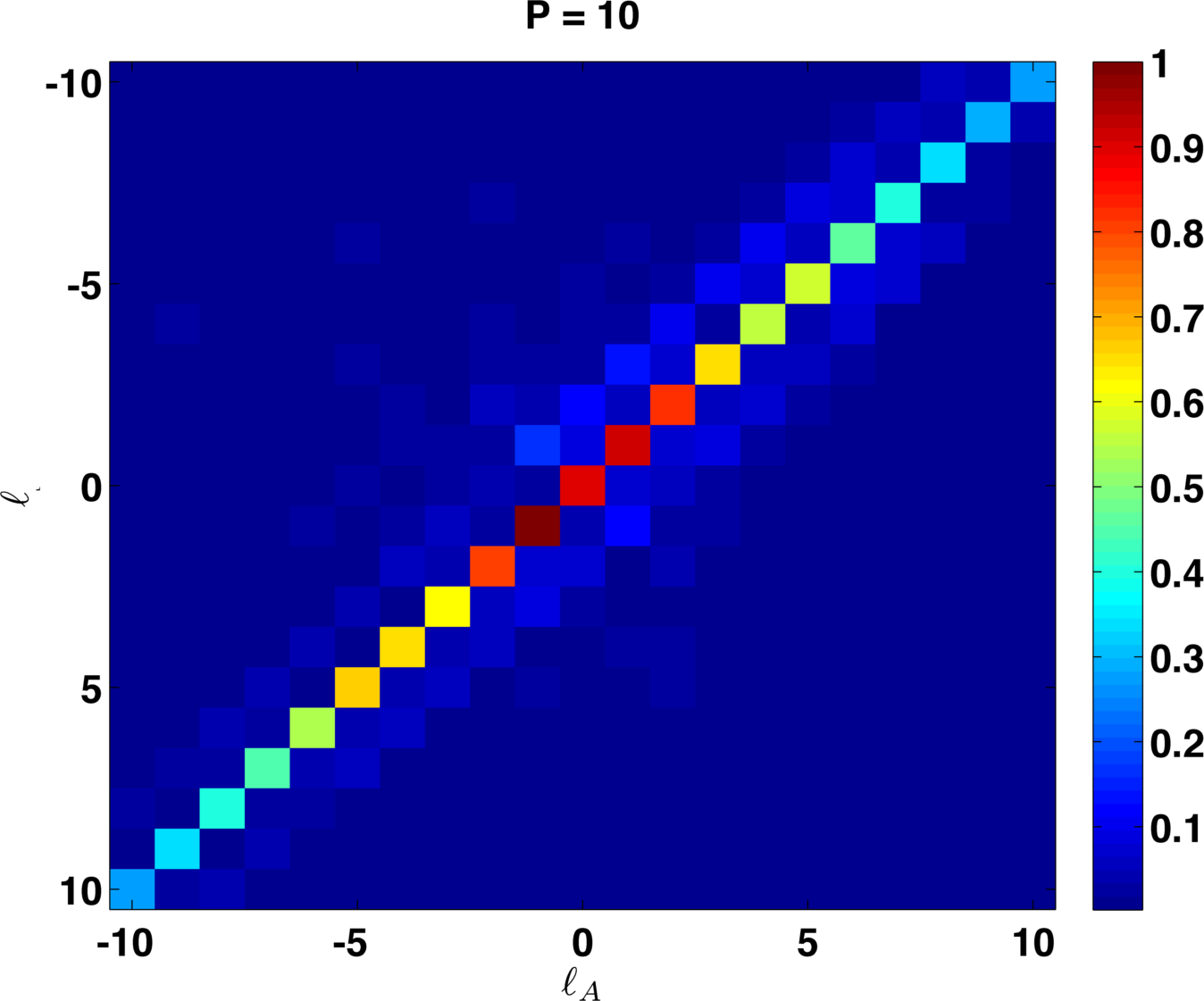}\tabularnewline
\end{tabular}

S1) Azimuthal quantum correlations at different $p_{s}=p_{i}$. We
see that the azimuthal quantum correlations are well preserved (i.e.,
that cross-correlations are small), also if the detected state involves
high radial quantum numbers. The axes in the graphs below are the
azimuthal quantum numbers $\ell_{A}$ (x-axis) and $\ell_{B}$ (y-axis).
The color coded pixels represent the coincidence counts, normalized
to unity.

\bigskip{}

\begin{tabular}{cc}
$\ell_{A}=-\ell_{B}=0$ & $\ell_{A}=-\ell_{B}=5$\tabularnewline
\includegraphics[width=0.45\columnwidth]{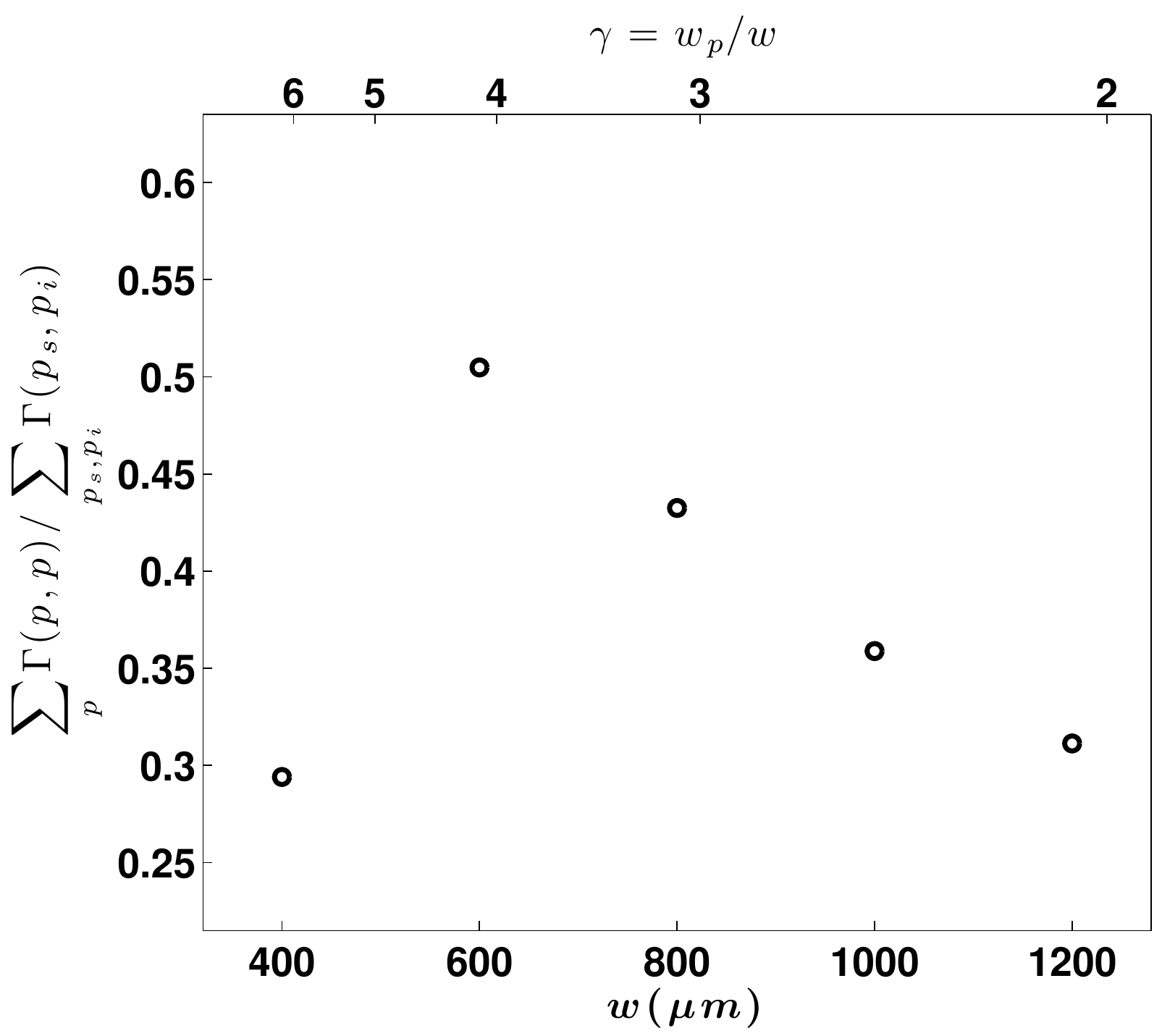} & \includegraphics[width=0.45\columnwidth]{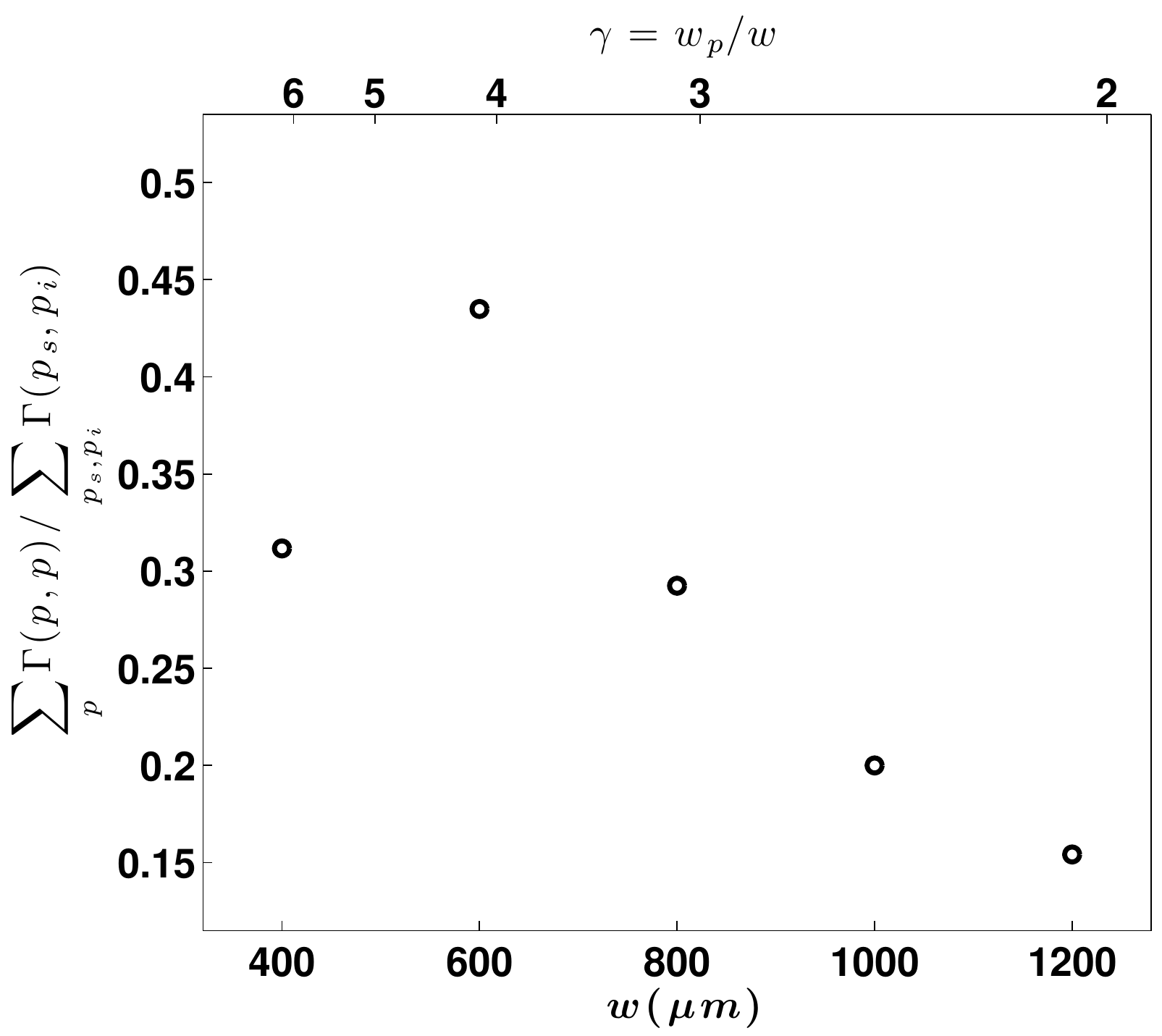}\tabularnewline
\end{tabular}

\bigskip{}

S2) Degree of correlation in $p$-space, similar to Fig.~3: Also
for higher azimuthal quantum numbers $\ell_{A}=-\ell_{B}$ we find
that, in general, by decreasing the detection mode waist, the degree
of correlation is increased. Only for very small detection mode waists
$w<600\,\mu$m we observe again (compare Fig.~3) an increase of undesired
cross-correlations.

\end{widetext} 
\end{document}